# SABRE Enhancement with Oscillating Pulse Sequences: Symmetry Reduces Robustness

Xiaoqing Li [a], Jacob R. Lindale[b], Shannon L. Eriksson[b, c], Warren S. Warren*[a, b, d]

[a] Prof. W. S. Warren, X. Li,
Department of Physics
Duke University
Durham, NC 27708 (USA)
E-mail: warren.warren@duke.edu

[b] Prof. W. S. Warren, Dr. J. R. Lindale, S. L. Eriksson
Department of Chemistry
Duke University
Durham, NC 27708 (USA)

[c] S. L. Eriksson
School of Medicine
Duke University
Durham, NC 27708 (USA)

[d] Prof. W. S. Warren
Department of Biomedical Engineering, and Radiology
Duke University
Durham, NC 27708 (USA)

**Abstract:** SABRE (Signal Amplification by Reversible Exchange) methods provide a simple, fast, and cost-effective method to hyperpolarize a wide variety of molecules in solution, and have been demonstrated with protons and, more recently, with heteronuclei (X-SABRE). The conventional analysis of the SABRE effect is based on level anti-crossings (LACs), which requires very low magnetic fields (~ 0.6µT) to achieve resonance and transfer spin order from the para-hydrogen to target heteronuclei. We have demonstrated in our recent study that the validity of LACs used in SABRE is very limited, so the maximum SABRE polarization predicted with LACs is not correct. Here, we present several oscillating pulse sequences that use magnetic fields far away from the resonance condition and can commonly triple the polarization. An analysis with average Hamiltonian theory indicates that the oscillating pulse, in effect, adjusts the J-couplings between hydrides and target nuclei and that a much weaker coupling produces maximum polarization. This theoretical treatment, combined with simulations and experiment, show substantial magnetization improvements relative to traditional X-SABRE methods. It also shows that, in contrast to most pulse sequence applications, waveforms with reduced time symmetry in the toggling frame make magnetization generation more robust to experimental imperfections.

## Introduction

Low sensitivity is an intrinsic limitation of nuclear magnetic resonance, because the energy difference caused by Zeeman splitting is normally much smaller than thermal energy, and the resultant equilibrium fractional magnetization is low (P ~ $10^{-5}$-$10^{-6}$). Hyperpolarization methods derive spin order from other sources and can create significantly higher magnetization. Three major methods have evolved over the last several decades: dissolution dynamic nuclear polarization (d-DNP)[1], which derives nuclear spin order from unpaired electrons, and spin exchange optical pumping (SEOP)[2], which derives it indirectly from circularly pumped optical transitions and hydrogenative para-hydrogen-induced polarization (PHIP)[3], which derives spin order from para-hydrogen, the singlet isomer of the $H_2$ molecule. Active research continues on all of these methods, in large part because they have obvious limitations. SEOP is restricted to a few noble gases, d-DNP needs high-cost hyperpolarization hardware and a long hyperpolarization time (often an hour or so for $^{13}$C and $^{15}$N), and PHIP requires a proper precursor molecule and catalyst.

More recently, a variety of methods have evolved which use reversible interactions of parahydrogen and a target molecule with an iridium catalyst, starting with the method known as Signal Amplification By Reversible Exchange (SABRE)[4-7]. Both parahydrogen and target substrate rapidly and reversibly exchange with sites on the catalyst metal center. In a low magnetic field (~6mT), J-couplings of the hydrides and protons on the bound species transfer spin order between them, and this makes it possible to spontaneously create excess magnetization on the target protons. In recent years, a variety of extended SABRE methods (X-SABRE) [8-15] have relaxed the experimental restrictions. For example, SABRE-SHEATH (Scheme 1) (Signal Amplification by Reversible Exchange in SHield Enables Alignment Transfer to Heteronuclei) has permitted direct targeting of heteronuclei ($^{15}$N, $^{13}$C, $^{19}$F, and $^{31}$P)[8, 16-20] with much longer $T_1$ values than $^1$H. In this case, the optimal magnetic field is about 0.6µT[21], so the experiments are generally done in a magnetic shield. Other X-SABRE methods have been adapted to transfer spin order from parahydrogen directly in a high field magnet[11-15].

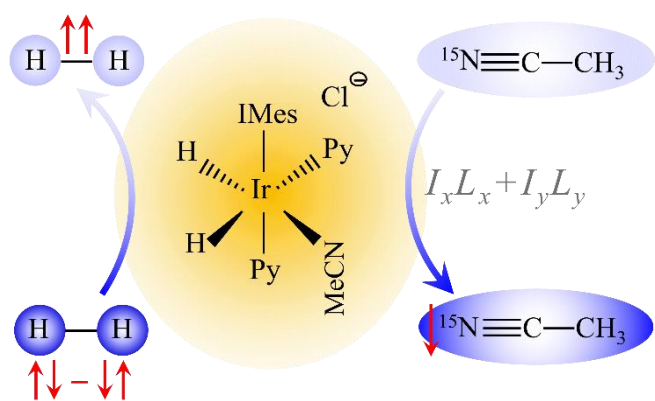

*Scheme 1.* Schematic representation of generation of hyperpolarized $^{15}$N labeled acetonitrile. Imes, Py and MeCN represent [1,3-bis(2,4,6-trimethyphenyl)-imidazoyl], pyridine ligands, and acetonitrile, respectively.

SABRE and X-SABRE are simpler, faster, and less expensive than commercially available hyperpolarization methods, and more general than PHIP. However, the amount of polarization produced at any one time is generally lower than with d-DNP or SEOP, although there is no fundamental reason why this must be true. We have recently shown that a big part of the reason is that the novel field regime for SABRE and X-SABRE (where even heteronuclear couplings can be readily interconverted between the strong and weak coupling limits), combined with the very complex exchange dynamics, imply that the method is theoretically underexplored; there are clearly better (but nonintuitive) approaches to creating polarization than a simple continuous field.

This paper focuses on a major advantage of operating in the low field regime: the ability to change the main magnetic field at will, much faster than any couplings, using very simple hardware. In particular, we explore the use of periodic field perturbations (Figure 1), with the goal of creating enhanced magnetization with low sensitivity to experimental imperfections such as field inhomogeneity. We obtain general insight from average

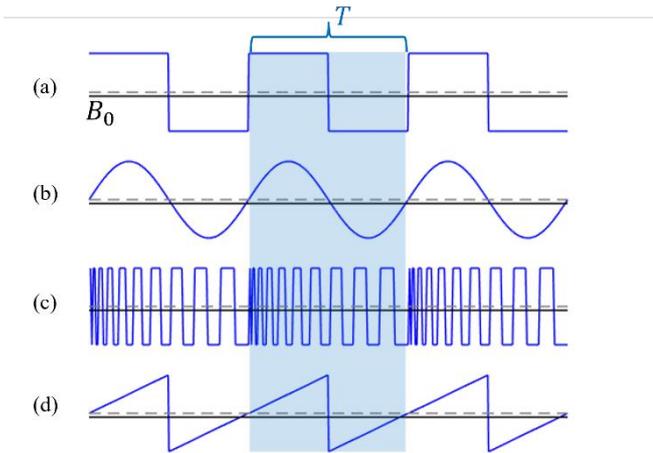

*Figure 1.* Pulse sequences for enhanced SABRE/X-SABRE excitation. (a) square pulse, (b) sine wave pulse, (c) chirped pulse, and (d) ramp pulse. The dashed grey line refers to zero magnetic field, and The black line is an offset field, $B_0$. As shown in this paper, all of these sequences are capable of producing signal enhancements (relative to a constant field) but the lower symmetry sequences (c) and (d) have practical advantages.

Hamiltonian theory[22, 23] and then do highly accurate calculations using an exact dissipative master equation treatment[24]. In all cases, optimal pulse sequences look nonintuitive and do not match continuous excitation, either in their peak or average field strength. Figure 2 shows one example; in this system, experimentally and in simulations the maximal magnetization (~5%) is generated with a continuous ~0.6μT field, but far larger magnetization (~18%) is produced by a correctly timed square wave offset from a zero average field by about one-fifth that value (~0.13μT). We will also show that, in contrast to most pulse sequence applications, waveforms with reduced time symmetry in the toggling frame (such as the last two sequences in Figure 1) make magnetization generation more robust to experimental imperfections.

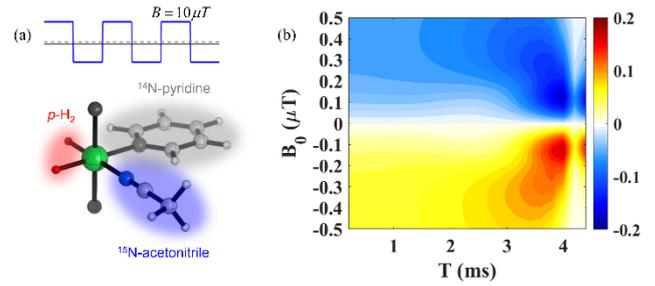

*Figure 2.* (a) Schematic representation of the pulse and the spin system used in simulation. The coupling strengh between the two hydrides is $J_{HH} = -8H$. It is often a good approximation to assume that the ligand is only coupled to one of the two hydrides, and the strength is $J_{HL} = -25 Hz$. (b) Final $^{15}$N polarization level simulated with the unbalanced square wave. The offset field $B_0$ is varied from -0.5μT to 0.5μT, and the pulse period T is scanned in the range of 0 to 4.4ms, while the pulse amplitude B is fixed at 10μT. Simulation parameters: 100% parahydrogen, $k_L = 24s^{-1}$, $k_H = 8s^{-1}$, $[catalyst]:[ligand] = 1:10$. A pulse with $B_0=\pm0.13μT$ and T=4ms yields the maximum ~18% polarization, which is much larger than the continuous wave counterpart ~5%.

## Theoretical Perspective on Oscillating Pulses

Each pulse sequence in Figure 1 consists of a low offset field (<1μT) and an oscillating pulse with alternating positive-negative amplitude (10-100μT). Every pulse is specified by three parameters--the offset field $B_0$, the oscillating field $B(t)$, and the pulse period T. We describe these sequences using average Hamiltonian theory[22, 23], which relies on the principle that under suitable conditions, the evolution of a spin system driven by a time-dependent external field can be described by the average effect of the field over one cycle of its oscillation. For the sake of comprehensiveness, we analyze both AA'B 3-spin system and AA'BB' 4-spin system (shown in SI section 3). The full Hamiltonian of the three-spin AA'B system is expressed in equation (1)

$$\hat{\mathcal{H}}_{prime}(t) = -(B_0 + B(t))(\gamma_H(\hat{I}_{1z} + \hat{I}_{2z}) + \gamma_L \hat{L}_z) + 2\pi J_{HH}(\vec{I}_1 \cdot \vec{I}_2) + 2\pi J_{HL}(\vec{I}_1 \cdot \vec{L}) \quad (1)$$

here in natural units ($\hbar = 1$). The spin operator of the target nuclei is labelled as $\vec{L}$, and the two hydride spins are $\vec{I}_1$ and $\vec{I}_2$. We can rearrange this Hamiltonian as:

$$\hat{\mathcal{H}}_{rearranged}(t) = -(B_0 + B(t))\gamma_H(\hat{I}_{1z} + \hat{I}_{2z} + \hat{L}_z) + (\Delta\omega_0 + \Delta\omega(t))\hat{L}_z + 2\pi J_{HH}(\vec{I}_1 \cdot \vec{I}_2) + 2\pi J_{HL}(\vec{I}_1 \cdot \vec{L}) \quad (2)$$

in which $\Delta\omega_0 = B_0(\gamma_H - \gamma_L)$ and $\Delta\omega(t) = B(t)(\gamma_H - \gamma_L)$ are the Larmor frequency difference between hydrides and the target nuclei caused by the offset field and the oscillating pulse respectively. The first term in equation (2) which is directly proportional to the z component of the total spin angular momentum can be ignored since it commutes with the rest of the Hamiltonian, giving a simplified Hamiltonian of the form

$$\hat{\mathcal{H}}(t) = (\Delta\omega_0 + \Delta\omega(t))\hat{L}_z + 2\pi J_{HH}(\vec{I}_1 \cdot \vec{I}_2) + 2\pi J_{HL}(\vec{I}_1 \cdot \vec{L}) \quad (3)$$

We use the only time-dependent term $\Delta\omega \hat{L}_z$ to create a toggling frame $U(t) = \exp(-i\hat{L}_z \int_0^t \Delta\omega(t') dt')$. The corresponding Hamiltonian in this toggling frame is equation (4).

$$\tilde{\mathcal{H}} = U\left(\Delta\omega_0 \hat{L}_z + 2\pi J_{HH}(\vec{I}_1 \cdot \vec{I}_2) + 2\pi J_{HL}(\vec{I}_1 \cdot \vec{L})\right)U^\dagger$$
$$= \Delta\omega_0 \hat{L}_z + 2\pi J_{HH}(\vec{I}_1 \cdot \vec{I}_2) + 2\pi J_{HL}\{\hat{I}_{1z}\hat{L}_z + M(t)(\hat{I}_{1x}\hat{L}_x + \hat{I}_{1y}\hat{L}_y) + N(t)(\hat{I}_{1x}\hat{L}_y - \hat{I}_{1y}\hat{L}_x)\} \quad (4)$$

in which $M(t) = \cos(\int_0^t \Delta\omega(t') dt')$, and $N(t) = \sin(\int_0^t \Delta\omega(t') dt')$. This toggling frame Hamiltonian has unveiled the physical picture of the pulse sequence. The role of the offset is to provide a small external magnetic field to the spins. The oscillating pulse then alters the form of the spin-spin interaction between the target nuclei and hydrides. The original flip-flop term, $\hat{I}_{1x}\hat{L}_x + \hat{I}_{1y}\hat{L}_y$, and the new interaction form, $\hat{I}_{1y}\hat{L}_x - \hat{I}_{1x}\hat{L}_y$, are tuned by the factors $M(t)$ and $N(t)$, respectively. These two terms connect the same states as the normal non-secular term, but with a $\pi/2$ phase shift of the off-diagonal operators, which will be clearer with the matrix form of the Hamiltonian shown later. When $T \to 0$, the oscillating pulse vanishes, and the system is recovered to continuous wave (CW) SABRE-SHEATH with $M = 1$ and $N = 0$.

For the square pulse (shown in Figure 1(a)) the zero-order average Hamiltonian is shown in equation (5).

$$\begin{aligned}\widetilde{\mathcal{H}}^{(0)} &= \frac{1}{T}\int_0^T \widetilde{\mathcal{H}}(t)dt \\ &= \Delta\omega_0 \hat{L}_z + 2\pi J_{HH}(\vec{I}_1 \cdot \vec{I}_2) \\ &+ 2\pi J_{HL}\{\hat{I}_{1z}\hat{L}_z + M_0(\hat{I}_{1x}\hat{L}_x + \hat{I}_{1y}\hat{L}_y) + N_0(\hat{I}_{1x}\hat{L}_y - \hat{I}_{1y}\hat{L}_x)\}\end{aligned} \quad (5)$$

where $M_0 = \frac{\sin(\theta)}{\theta}$, $N_0 = \frac{1-\cos(\theta)}{\theta}$, and $\theta = \Delta\omega T/2$ representing the rotation angle in half a period is a function of pulse amplitude and pulse period. When $\theta$ is an integer multiple of $2\pi$, both $M_0$ and $N_0$ go to zero, which means the coupling between hydrides and the target nuclei disappear at this situation, and no spin order transfer could take place.

A matrix expression of this zero-order Hamiltonian is powerful for providing physical insight. The basis used to express the matrix of the AA'B system is a singlet-triplet basis for the AA' pair and the Zeeman basis for the B spin. Equation (6) gives the two 3×3 subspaces of the zero-order Hamiltonian, which indicate that $M_0 \pm iN_0$ alter the interaction between the spin up states $\alpha_L$ and spin down states $\beta_L$ of the target nuclei, and that the interaction strength only depends on its magnitude, $\sqrt{M_0^2 + N_0^2}$.

$$\begin{array}{c} \begin{array}{ccc} T_H^+\beta_L & T_H^0\alpha_L & S_H^0\alpha_L \end{array} \\ \begin{array}{c}T_H^+\beta_L \\ T_H^0\alpha_L \\ S_H^0\alpha_L\end{array} \left( \begin{array}{ccc} \frac{\pi(J_{HH}-J_{HL})-\Delta\omega_0}{2} & \frac{\pi J_{HL}}{\sqrt{2}}(M_0+iN_0) & \frac{-\pi J_{HL}}{\sqrt{2}}(M_0+iN_0) \\ \frac{\pi J_{HL}}{\sqrt{2}}(M_0-iN_0) & \frac{\pi J_{HH}+\Delta\omega_0}{2} & \frac{\pi J_{HL}}{2} \\ \frac{-\pi J_{HL}}{\sqrt{2}}(M_0-iN_0) & \frac{\pi J_{HL}}{2} & \frac{\Delta\omega_0-3\pi J_{HH}}{2} \end{array} \right) \end{array} \quad (6)$$

$$\begin{array}{c} \begin{array}{ccc} T_H^-\alpha_L & T_H^0\beta_L & S_H^0\beta_L \end{array} \\ \begin{array}{c}T_H^-\alpha_L \\ T_H^0\beta_L \\ S_H^0\beta_L\end{array} \left( \begin{array}{ccc} \frac{\pi(J_{HH}-J_{HL})+\Delta\omega_0}{2} & \frac{\pi J_{HL}}{\sqrt{2}}(M_0-iN_0) & \frac{\pi J_{HL}}{\sqrt{2}}(M_0-iN_0) \\ \frac{\pi J_{HL}}{\sqrt{2}}(M_0+iN_0) & \frac{\pi J_{HH}-\Delta\omega_0}{2} & \frac{-\pi J_{HL}}{2} \\ \frac{\pi J_{HL}}{\sqrt{2}}(M_0+iN_0) & \frac{-\pi J_{HL}}{2} & \frac{-3\pi J_{HH}-\Delta\omega_0}{2} \end{array} \right) \end{array}$$

While it would be possible to explicitly calculate higher order terms (S1 section 1) direct numerical evaluation of the full effective Hamiltonian gives a better comparison to test the validity. Each cycle of the square-pulse can be broken down into two constant fields, $B_0 + B$ and $B_0 - B$, in time sequence. Labeling the corresponding time-independent Hamiltonians as $\mathcal{H}_+$ and $\mathcal{H}_-$, the propagator of this spin system is expressed as $U = \exp(-i\mathcal{H}_-T/2)\exp(-i\mathcal{H}_+T/2) = \exp(-i\overline{\mathcal{H}}T)$. Extracting the average Hamiltonian by an expression such as $\overline{\mathcal{H}} = i \log(U)/T$ can be done numerically by diagonalizing $U$ to a matrix $e^{i\Lambda}$, then taking the log of each eigenvalue (which will always have magnitude 1), $U = V e^{i\Lambda} V^\dagger$; $\log U = V(i\Lambda) V^\dagger$, but this leads to a well-known ambiguity as the phase is only determined modulo $2\pi$. This ambiguity is avoided by multiplying $J_{HH}$ and $J_{HL}$ by a scale factor $\alpha$, calculating $\overline{\mathcal{H}}$ in the limit of very small $\alpha$, and correcting for $2\pi$ phase jumps as $\alpha$ is increased to 1. This approach shows that the zero-order expansion is already a good approximation of the average Hamiltonian (Figure 3) so that higher order approximation can be neglected. $M_0$, $N_0$ and $\sqrt{M_0^2 + N_0^2}$ (solid curves) calculated with the zero-order average Hamiltonian are in great agreement with their numerical counterparts (dashed curves). The coupling magnitude $\sqrt{M_0^2 + N_0^2}$ vanishes at $\theta = 2n\pi$, which agrees with the analytical expressions of $M_0$ and $N_0$. $\sqrt{M_0^2 + N_0^2}$ attenuates as $\theta$ increases, in that it cannot fully recover to the previous maximum. The value of $\sqrt{M_0^2 + N_0^2}$ is in the range of 0 to 1, which is obvious from the formula of $M(t)$ and $N(t)$ because they are conjugate trigonometric functions. In other words, the coupling between hydrides and the target nuclei can only be attenuated instead of being increased.

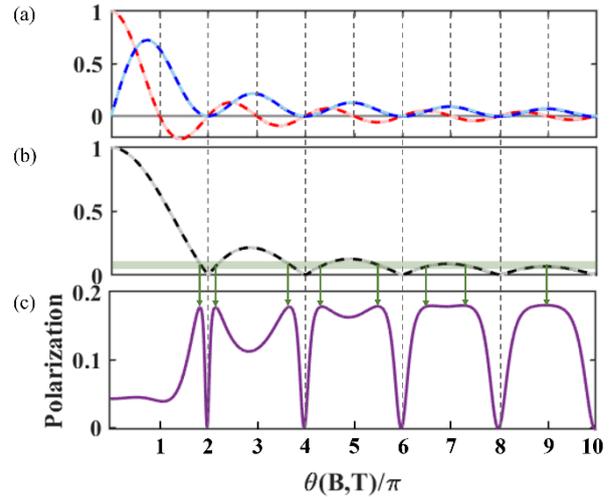

*Figure 3.* (a) Curves of M and N as a function of $\theta/\pi$. The solid lines (underlying bright red and bright blue curves) are the zero-order theoretical approximation of M and N, namely, $M_0$ and $N_0$, and the dashed lines (light red and blue) are the exact values of M and N calculated with numerical method. (b) shows how the value of $\sqrt{M^2 + N^2}$ varies with $\theta/\pi$. Likewise, the solid curve (underlying black) and the dashed line (grey) correspond to the theoretical approximation and the numerical result, respectively. The horizontal green bar indicates the optimal value of $\sqrt{M^2 + N^2}$. (c) Numerically simulated $^{15}$N polarization of a 3-spin system with the pulse amplitude being fixed at 10μT and the offset being maintained at -0.13μT. The green arrows mark the maximum signals and their corresponding value of $\sqrt{M^2 + N^2}$.

Unexpectedly, a diminished coupling strength yields much higher polarization. Figure 3(b) and (c) gives the relationship between the final polarization level and the value of $\sqrt{M_0^2 + N_0^2}$ with the offset field $B_0$ fixed at $-0.13\mu T$. The polarization is numerically simulated with the DMEx method[24], and the dependence on $\theta$ indicates that when the interaction strength reduces to $\sqrt{M_0^2 + N_0^2} \sim 0.066$ (shown with a horizontal green bar in figure 3(b)), polarization is maximized. This unbalanced square wave indeed yields a large increase in signal. However, the optimal interaction strength $\sqrt{M_0^2 + N_0^2} \sim 0.066$ is very close to zero, where the polarization oscillations in Figure 3 imply, hence the large signals are not robust to imperfections of the pulse sequence. This issue could be avoided if $\sqrt{M_0^2 + N_0^2}$ reduces gradually and does not periodically go to zero. To find a pulse sequence with this behavior of $\sqrt{M_0^2 + N_0^2}$, more complex wave forms must be considered.

Figure 4(a) shows that $M_0$, $N_0$, and $\sqrt{M_0^2 + N_0^2}$ of a sine wave have similar behaviors as they do for a square pulse. $M_0$ and $N_0$ in the zero-order average Hamiltonian are $M_0 = \frac{1}{T}\int_0^T \cos\{\frac{\Delta\omega T}{2\pi}(1-\cos\frac{2\pi t}{T})\}dt$ and $N_0 = \frac{1}{T}\int_0^T \sin\{\frac{\Delta\omega T}{2\pi}(1-\cos\frac{2\pi t}{T})\}dt$, and they periodically vanish at the same time. For all these sequences, the toggling frame is symmetric about the center of each repeating interval (for example, the negative part of the square wave retraces back along the same trajectory followed in the positive part). In addition, both the square and sine wave have a symmetric trajectory with respect to the center of each half cycle (depicted here as green and orange). This additional symmetry intuitively would be expected to produce cleaner pumping dynamics. However, it turns out that this symmetry causes $M_0$ and $N_0$ to commonly go to zero simultaneously for the same value of the cycle length--a problem which is avoided by waveforms with lower symmetry (Figures 1(c-d) and 4(b-c)).

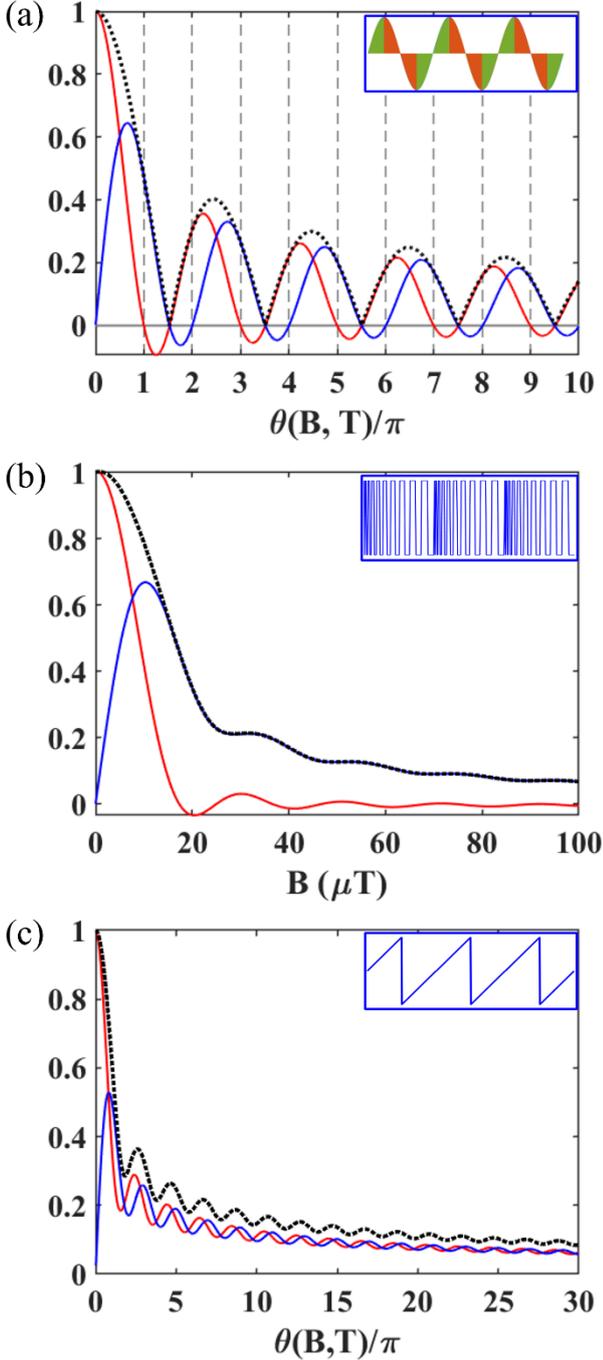

**Figure 4.** Depiction of how $M_0$ (red curve), $N_0$ (blue curve) and $\sqrt{M_0^2 + N_0^2}$ (black dotted curve) vary as a function of the rotation angle in half a period for a sine wave (a), a chirped square wave (b) and a sawtooth wave (c). We maintain the offset field at $-0.13\mu T$. As in Figure 3, frequent zero-crossings impose a serious constraint for the sine wave case on the usable fields, particularly if inhomogeneity is present, but not for the other cases.

To understand this effect, note that for an arbitrary oscillating pulse shape, $M_0$ and $N_0$ are given by

$$\begin{cases} M_0 = \frac{1}{T}\int_0^T M(t)\,dt = \frac{1}{T}\int_0^T \cos\left(\int_0^t \Delta\omega(t')dt'\right)dt \\ N_0 = \frac{1}{T}\int_0^T N(t)\,dt = \frac{1}{T}\int_0^T \sin\left(\int_0^t \Delta\omega(t')dt'\right)dt \end{cases} \quad (7)$$

which are integrals of the cosine and sine function of the instantaneous angle $\int_0^t \Delta\omega(t')dt'$, or in other words, integrals of the projections of a rotating unit vector on the x-axis and y-axis, respectively, in a rectangular coordinate system (Figure 5). The trajectory of the unit vector in the first quarter of a period is plotted as the green area; in the second quarter the vector keeps moving clockwise but with inversely changing speed. Then the vector retraces its steps and completes a full period. Note that in both Figure 3(a) and Figure 4(a), zero crossings of $M_0$ occur with positive lobe areas of $(2n + 1)\pi$ and zero crossings of $N_0$ occur at areas of $2n\pi$. These are simple symmetry effects, made clear by plotting the instantaneous values of $M(t)$ and $N(t)$ in the unit plane (Figure 5). In general, the trajectory is symmetric about its midway point $\theta/2$, which fixes the ratio $N_0/M_0 = \tan(\theta/2)$. For odd multiples of $\pi$ $M_0 = 0$, and for even multiples $N_0 = 0$. Other than for those values of $\theta$, the zero crossings of $M_0$ and $N_0$ depend on the details of the waveform. However, because the ratio between $M_0$ and $N_0$ is fixed, zeroes at any $\theta$ value must coincide, giving zero efficiency for generating polarization. Thus any waveform which has this symmetry (or can be given this symmetry by a time shift and a prepulse, such as an unbalanced square wave with different amplitudes but the same area in the positive and negative lobes), the enforced simultaneity of $M_0$ and $N_0$ zero crossings creates a highly structured pumping profile. In contrast, the lower symmetry in Figures 4b-c avoids simultaneous zeroes.

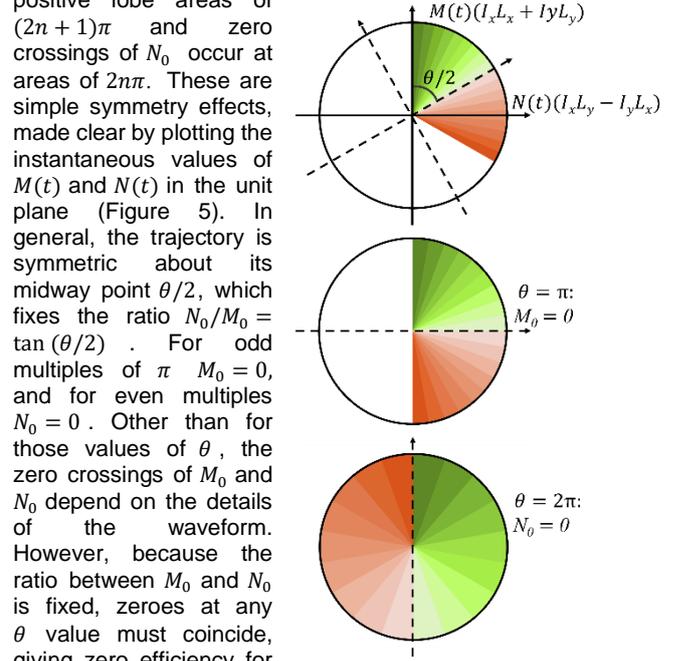

**Figure 5.** For many common waveforms (such as a sine or square wave) the time-dependent rotation between (xx+yy) and (xy-yx) terms is antisymmetric about the midpoint, and each lobe is symmetric. In this case, the relative values of the average coefficients of these two terms ($M_0$ and $N_0$ respectively) is constrained by symmetry to be along an axis with half the rotation angle of each lobe. Except for the special cases $\theta=n\pi$, this means that zeroes in $M_0$ and $N_0$ must coincide, creating values which generate no polarization.

Figure 1(c) and 4(b) show a chirped pulse with evenly growing wavelength in each period, $\tau_0$, $\tau_0 + \Delta\tau$, ... $\tau_0 + m\Delta\tau$, and $\sum_{j=0}^m (\tau_0 + j\Delta\tau) = T$. The chirped pulse we use here has $\tau_0 = \Delta\tau = 0.2 ms$. $\sqrt{M_0^2 + N_0^2}$ of this pulse is close to the optimal value 0.066 when the pulse amplitude B is larger than $40\mu T$, accordingly the resulting experimental robustness to the pulse amplitude is indeed improved. We finally try an asymmetric pulse shape, a ramp pulse (Figure 1(d)), which turns out to be robust to both the pulse period and the field strength. The analytical solution of a ramp pulse is similar to equation (5), excecpt that $M_0$ and $N_0$ are replaced by $M_0 = \frac{2}{\sqrt{\Delta\omega T}}\int_0^{\sqrt{\Delta\omega T}/2} \cos x^2 dx$, and $N_0 = \frac{2}{\sqrt{\Delta\omega T}}\int_0^{\sqrt{\Delta\omega T}/2} \sin x^2 dx$. Unlike the square pulse and sine wave whose effective $J_{HL}$ coupling can be fully averaged out when M and N become zero at the same time. The ramp wave successfully avoids zero points. Because the coupling strength $\sqrt{M_0^2 + N_0^2}$ gradually approaches zero as the pulse period or pulse amplitude increase (namely $\theta(B,T)$ increases), in a wide period and amplitude domain it always stays close to the optimal value 0.066, indicated in Figure 4(c).

For continuous wave excitation in SABRE/X-SABRE, we recently pointed out[25] that the level anticrossing[11, 26-29] condition does not even give qualitatively correct predictions except for very small $J_{HL}$ couplings and very slow exchange. For example, in the case of the AA'B system shown in Figure 2(a) with $J_{HH} = -8Hz$, and $J_{HL} = -25Hz$ in a continuous low magnetic field, the LAC occurs at $\pm 0.04\mu T$ which is far from the experimental optimum $\pm 0.6\mu T$. The failure of the level anticrossing condition here is mainly because it oversimplifies a $3 \times 3$ (or larger) subspace to a $2 \times 2$ space and usually cannot accurately account for the dynamics of the original system. Interestingly, though, the LAC condition becomes more relevant for oscillating pulse SABRE/X-SABRE, because the oscillating pulses reduce the off-diagonal elements in equation (6) but not the diagonal ones, thus improving the separation from unwanted states.

Figure 6 depicts how the energy levels of the first subspace in equation (6) vary with the offset field. The subfigure 6(a) relates to $J_{HL} = 0$, and no LACs occur because there is no interaction between the states. While Figure 6(b) refers to the case of optimal interaction $J_{HL} = -25Hz$ and $\sqrt{M_0^2 + N_0^2} \sim 0.066$, in which the circled LAC is in great agreement with the optimal offset field $\pm 0.13\mu T$.

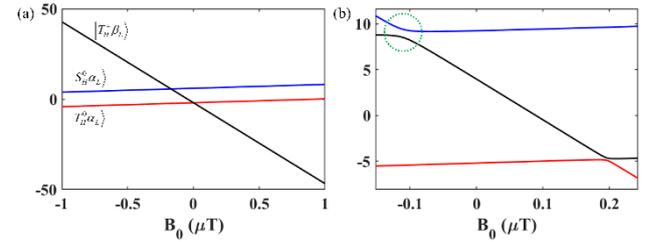

**Figure 6.** Eigenvalues as a function of the offset field. (a) corresponds to the case of $J_{HL} = 0$, while (b) refers to the case of optimal interaction $J_{HL} = -25$Hz and $\sqrt{M_0^2 + N_0^2} \sim 0.066$.

## Results and Discussion

In this section, we verify the analytical results above with both simulations and experiments. All simulations are done with the DMEx method[24] which is a recently developed numerical modeling approach for exchanging systems and has shown robust agreement with experimental results. All oscillating pulse SABRE-SHEATH experiments were performed by bubbling 43% parahydrogen through a methanol-d4 solution under 7 bars of pressure at room temperature. The SABRE sample used here was prepared by adding 15N-acetonitrile (50mM), natural-abundance 14N-pyridine (25mM), and the catalyst precursor [IrCl(COD)(IMes)] (4.4 mM) into 500μL deuterated methanol solvent. The catalyst was first activated by bubbling hydrogen gas through the SABRE sample for 30-60min to generate SABRE complex [IrH$_2$($^{15}$N-Py)$_2$(IMes)]$^+$($^{15}$N-py = $^{15}$N-pyridine). The whole sample was then bubbled for 60s inside a solenoid coil within a μ-metal magnetic shield in which the polarization transfer occurs. We connect the solenoid coil to a function generator to create different oscillating fields inside the coil. Finally, the sample was manually transferred (1-2s) to a 1 Tesla $^{15}$N Magritek NMR spectrometer for detection.

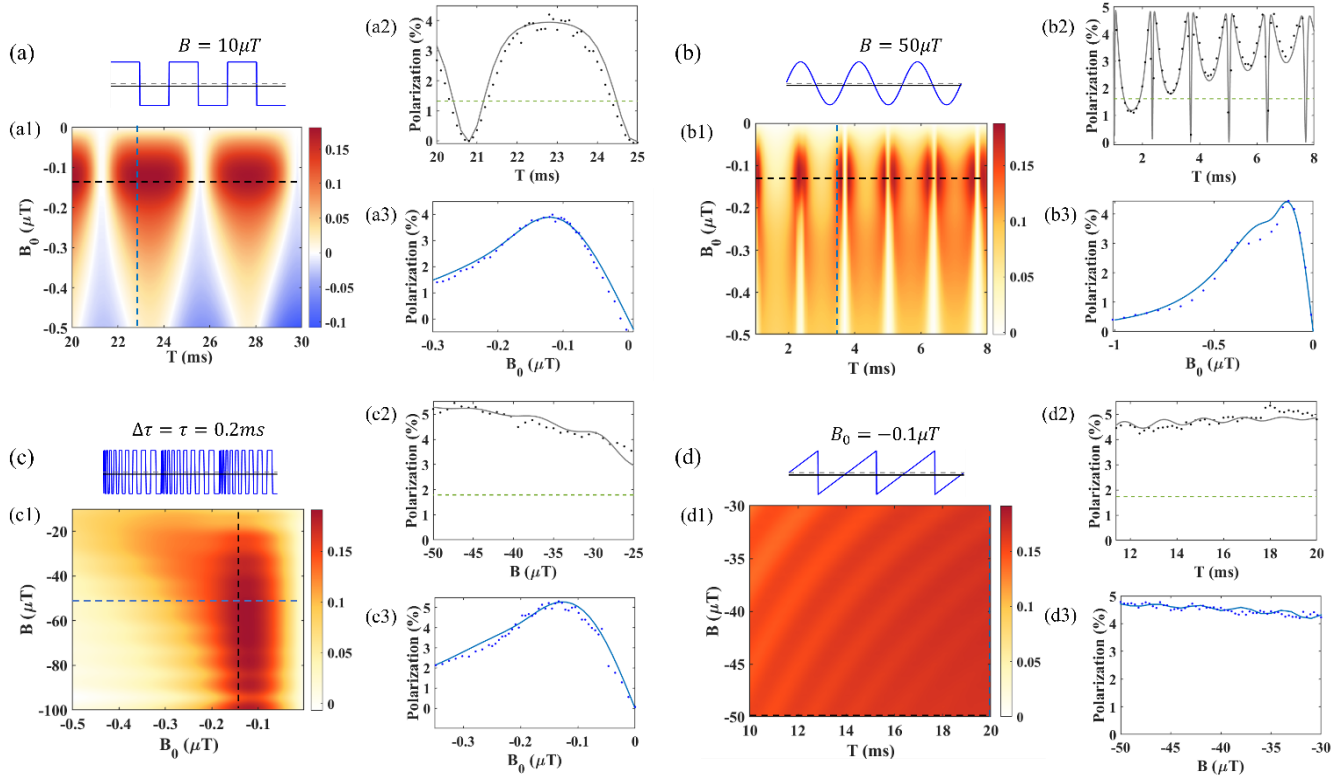

**Figure 7.** Experimental validation of theoretical predictions. (a) Square pulse (a1) Theoretical effects of a square pulse sequence with a fixed amplitude $B = 10\mu T$ and various pulse periods and offset fields. (a2) Comparison of theoretical calculations with experimental data holding $B_0 = -0.13\mu T$ and varying T. (a2) Comparison of theoretical calculations with experimental data holding $T = 22.7ms$ and varying $B_0$. (b) Sine wave (b1) Theoretical effects of a sine wave sequence with a fixed amplitude $B = 50\mu T$ and changing pulse periods and offset fields. (b2) Comparison of theoretical calculations with experimental data holding $B_0 = -0.13\mu T$ and varying T. (b3) Comparison of theoretical calculations with experimental data holding $T = 3.5ms$ and varying $B_0$. (c) Chirped pulse (c1) Theoretical effects of a chirped pulse with evenly increasing pulse length from 0.2ms to 2ms in a step of 0.2ms. The offset and the pulse amplitude are both scanned. (c2) Comparison of theoretical calculations with experimental data holding $B_0 = -0.13\mu T$ and varying B. (c3) Comparison of theoretical calculations with experimental data holding $B = 50\mu T$ and varying $B_0$. (d) Ramp pulse (d1) Theoretical effects of a ramp pulse with a fixed offset field $B_0 = -0.1\mu T$. (d2) Comparison of theoretical calculations with experimental data holding $B = 50\mu T$ and varying T. (d3) Comparison of theoretical calculations with experimental data holding $T = 20ms$ and varying B. The green dashed lines in each subfigure (2) refer to the maximum polarization obtained with the CW SABRE-SHEATH method using the same sample. Simulation parameters: 100% parahydrogen, $k_L = 24s^{-1}$, $k_H = 8s^{-1}$, [catalyst]: [ligand] = 1: 10, $J_{HH} = -8Hz$, $J_{HL} = -25Hz$. The maximum polarization for a continuous field experiment (with 43% p-H2 as used experimentally) is marked by the dashed green lines (~1.6%).

We start with the square pulse sequence (Figure 7(a)). The 3D plot Figure 7(a1) shows how the final polarization varies with the pulse period and the offset field while the pulse amplitude is fixed at $10\mu T$. The optimal offset field here is around $\sim 0.13\mu T$ rather than $\sim 0.6\mu T$ in the CW SABRE-SHEATH case even though the oscillating pulse has nothing to do with the Zeeman terms of the spin system. Besides, at the points $T = 20.8ms$, and $T = 25\ ms$ which make $\theta = 2n\pi$, no polarization is produced. All the analysis based on theory is in accordance with the experimental results. Figure 7(b) shows the polarization of a sine wave pulse as a function of both the pulse period and the offset field while the pulse amplitude is maintained at 50μT. We varied either the pulse period ($1ms \leq T \leq 8ms$, $B_0 = -0.13\mu T$) in Figure 7(b2), or the offset field ($-1\mu T \leq B_0 \leq 0\mu T$, $T = 3.5ms$) in Figure 7(b3). In agreement with the theoretical predictions, the final polarization periodically reduces to zero at a frequency corresponding to $\Delta\omega$, and the optimal pulse periods come close to zero polarization points. The optimal offset field is also shifted from $\sim 0.6\mu T$ to $\sim 0.13\mu T$. The result of a chirped pulse is displayed in Figure 7(c). Provided the offset is maintained in the range from $-0.18\mu T$ to $-0.08\mu T$, the polarization is robust to all pulse amplitudes larger than $40\mu T$ and stays within the range from 16.5% to 19%. The last oscillating pulse, ramp pulse, is shown in Figure 7(d). By varying the pulse period and amplitude while fixing the offset field at $-0.1\mu T$, the simulation result, Figure 7(d1), clearly shows that in a fairly wide range of both the pulse period and the pulse amplitude the final polarization always stays close to the maximum, which has been confirmed by experiments, Figure 7(d2) and 7(d3). By varying either the pulse period ($11ms \leq T \leq 20ms$, $B = 50\mu T$, $B_0 = -0.1\mu T$) or the pulse amplitude ($30\mu T \leq B \leq 50\mu T$, $T = 20ms$, $B_0 = -0.1\mu T$), we always obtained almost maximum polarization. The signal enhancement of these oscillation pulse sequences is as high as 300%.

## Robustness to Exchange Rate

In this section, by using DMEx simulation method[24], we demonstrate that oscillating pulse SABRE-SHEATH is robust to variations of exchange rate. We use the same square pulse in figure 2 and locate the maximum polarizations of four different sets of exchange rates. In the case of low substrate exchange rate (Figure 8(a) and (b)), the optimal condition stays around $B_0 = -0.11\mu T$, $T = 4.1ms$. However, as the exchange rate of the substrate goes up, the optimal condition shifts in the direction that the offset field increases while the pulse period decreases. Here we give an explanation with the quantum dynamics of SABRE. Figure 9 shows the polarization of the target nuclei as a function of the quantum evolution time. Both chemical exchange and relaxation processes are neglected. The four curves in Figure 9 depicts the quantum dynamics using optimal parameters $B_0$ and $T$ for the four cases of figure 8, respectively. The red and blue curves clearly show that under the condition $B_0 = -0.11\mu T$ and $T = 4.1ms$, the polarization of the target nuclei could be raised up to 80%. However, when the chemical exchange processes are included, the quantum evolution is interrupted, which means the spin order transfer is interrupted. On one hand, if the lifetime of the SABRE complex is long, for example $k_L = 1s^{-1}$ or $k_L = 10s^{-1}$, there is enough time that the target nuclei could be polarized before they dissociate from SABRE catalyst. On the other hand, during a short lifetime, say $k_L = 50s^{-1}$ or $k_L = 100s^{-1}$, the large degree of spin order transfer cannot be completed. A smaller but faster polarization transfer process works better (the green and black curves). As pulse period T decreases, the value of $\sqrt{M^2 + N^2}$ grows, and the coupling between the hydrides and the target nuclei increases accordingly, which causes the spin order transfer faster (shown in Figure 9(b)). For a fixed pulse period T, there exists an optimal offset field which maximizes the polarization of the substrate. In conclusion, maximizing SABRE polarization is a trade-off between the speed of spin order transfer and the largest transfer degree. Usually, the exchange rate $k_L$ is in the range from $5s^{-1}$ to $50s^{-1}$, and the corresponding optimal condition remains near $B_0 = -0.11\mu T$ and $T = 4.1ms$, (see Figure 8(b) and (c)). Even if the optimal condition shifts, for example, Figure 8(d), the polarization (1.4%) generated by an oscillating pulse with $B_0 = -0.11\mu T$ and $T = 4.1ms$ is still larger than the maximum signal (0.2%) produced by CW SABRE-SHEATH method. This robustness to exchange rate is also extended to AA'BB' system and a positive $J_{HH}$ situation (SI section 5).

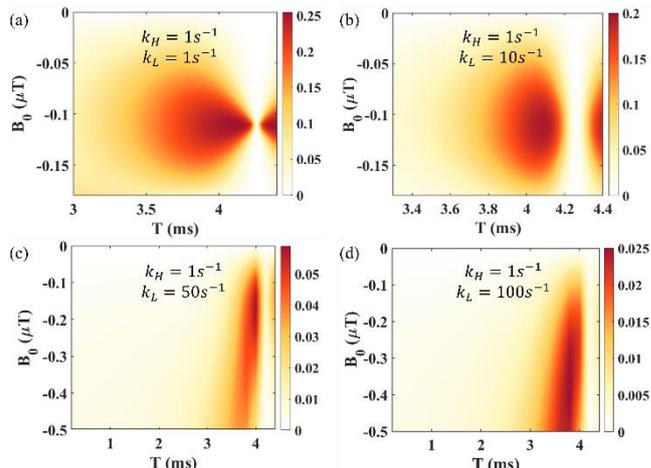

**Figure 8.** Different optimal polarization conditions caused by different exchange rates of substrate. (a) $k_H = 1s^{-1}$, $k_L = 1s^{-1}$, the optimal field is -0.11μT, and the optimal period is 4.12ms. (b) $k_H = 1s^{-1}$, $k_L = 10s^{-1}$, the optimal field is -0.11μT, and the optimal period is 4.08ms. (c) $k_H = 1s^{-1}$, $k_L = 50s^{-1}$, the optimal field is -0.17μT, and the optimal period is 4.0ms. (d) $k_H = 1s^{-1}$, $k_L = 100s^{-1}$, the optimal field is -0.37μT, and the optimal period is 3.8ms. The four subplots have inconsistent color scales in order to show off more detail.

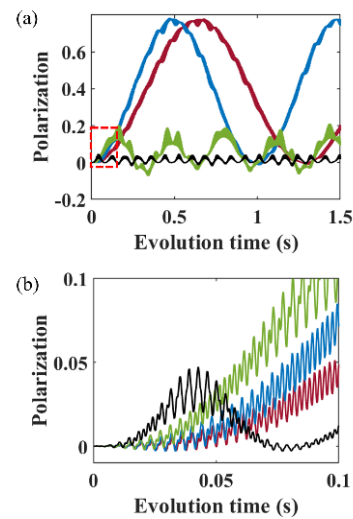

**Figure 9.** (a) Polarization generated by quantum evolution of the 3-spin system. red curve $B_0 = -0.11\mu T$, $T = 4.12ms$; blue curve $B_0 = -0.11\mu T$, $T = 4.08ms$; green curve $B_0 = -0.17\mu T$, $T = 4.0ms$; black curve $B_0 = -0.31\mu T$, $T = 3.8ms$. (b) A zoomed-in version of the red box in figure (a). As the period goes down the spin order transfer process speeds up.

## Conclusions

We have shown, in both experiments and simulations, that a variety of oscillating magnetic fields can significantly improve SABRE-SHEATH hyperpolarization, relative to a continuous field or even one pulsed to optimize polarization transfer. In effect, the interactions between hydrides and target nuclei are adjustable by tuning the pulse amplitude, pulse period, and pulse shape. As described above, we improve the robustness to experimental imperfections by exploring different pulse shapes. Finally, it turns out that a pulse shape with reduced symmetry, such as a ramp wave, generates significant improvements in achievable polarization and is robust to experimental imperfection.

In this work, the largest polarization yielded with 43% parahydrogen is ~5.3%; this would correspond to 22% polarization with 100% parahydrogen. In our simulations, the maximum polarization generated by SABRE-SHEATH method is as high as 80% (with 100% parahydrogen). However, in reality, polarization is limited by the low accessibility of parahydrogen or slow refresh rate of the dissolved hydrogen in the SABRE solvent. This gap between principle and reality indicates that SABRE retains room for improvement. Still, our approach is technologically simple, capable of producing significant enhancements, and suggests there is still room for further pulse sequence exploration.

## Acknowledgements

This work was supported by the National Science Foundation under grant CHE-2003109

**Keywords:** hyperpolarization • NMR spectroscopy • SABRE • polarization transfer • oscillating pulse • average Hamiltonian

# SABRE Enhancement with Oscillating Pulse Sequences: Symmetry Reduces Robustness

Xiaoqing Li*, Jacob R. Lindale, Shannon L. Eriksson, Warren S. Warren*

**Abstract:** SABRE (Signal Amplification by Reversible Exchange) methods provide a simple, fast, and cost-effective method to hyperpolarize a wide variety of molecules in solution, and have been demonstrated with protons and, more recently, with heteronuclei (X-SABRE). The conventional analysis of the SABRE effect is based on level anti-crossings (LACs), which requires very low magnetic fields (~ 0.6µT) to achieve resonance and transfer spin order from the para-hydrogen to target heteronuclei. We have demonstrated in our recent study that the validity of LACs used in SABRE is very limited, so the maximum SABRE polarization predicted with LACs is not correct. Here, we present several oscillating pulse sequences that use magnetic fields far away from the resonance condition and can commonly triple the polarization. An analysis with average Hamiltonian theory indicates that the oscillating pulse, in effect, adjusts the J-couplings between hydrides and target nuclei and that a much weaker coupling produces maximum polarization. This theoretical treatment, combined with simulations and experiment, show substantial magnetization improvements relative to traditional X-SABRE methods. It also shows that, in contrast to most pulse sequence applications, waveforms with reduced time symmetry in the toggling frame make magnetization generation more robust to experimental imperfections.

# SUPPORTING INFORMATION

## Table of Contents

## Higher-order Average Hamiltonian of AA'B System

In the theoretical section of the article, we ignore the higher-order approximation of the average Hamiltonian[1,2] and only discuss the zero-order Magnus expansion. Here, we will correct this omission. The first and second orders of the Magnus expansion are given by

$$\widetilde{\mathcal{H}}^{(1)} = \frac{1}{2iT}\int_0^T dt_1 \int_0^{t_1} dt_2\, [\widetilde{\mathcal{H}}(t_1),\widetilde{\mathcal{H}}(t_2)] \tag{S1}$$

$$\widetilde{\mathcal{H}}^{(2)} = \frac{1}{6T}\int_0^T dt_1 \int_0^{t_1} dt_2 \int_0^{t_2} dt_3\, \{[\widetilde{\mathcal{H}}(t_1),[\widetilde{\mathcal{H}}(t_2),\widetilde{\mathcal{H}}(t_3)]] + [[\widetilde{\mathcal{H}}(t_1),\widetilde{\mathcal{H}}(t_2)],\widetilde{\mathcal{H}}(t_3)]\} \tag{S2}$$

In the square pulse and the sine wave cases, since $\widetilde{\mathcal{H}}(t)$ is symmetric, $\widetilde{\mathcal{H}}(t) = \widetilde{\mathcal{H}}(T-t)$ for any $0 \le t \le T$, all odd order Magnus expansions are zero. Here, we give the commutation relation of $[\widetilde{\mathcal{H}}(t_1),\widetilde{\mathcal{H}}(t_2)]$ and $[[\widetilde{\mathcal{H}}(t_1),\widetilde{\mathcal{H}}(t_2)],\widetilde{\mathcal{H}}(t_3)]$, based on which $\widetilde{\mathcal{H}}^{(1)}$ and $\widetilde{\mathcal{H}}^{(2)}$ can be calculated with integral computations.

$$[\widetilde{\mathcal{H}}(t_1),\widetilde{\mathcal{H}}(t_2)] = \frac{i}{2}\Delta\omega_0 J_{HL}\{(M(t_1)-M(t_2))(\hat{I}_{1x}\hat{L}_y - \hat{I}_{1y}\hat{L}_x) + (N(t_1)-N(t_2))(\hat{I}_{1x}\hat{L}_x + \hat{I}_{1y}\hat{L}_y)\} \tag{S3}$$
$$+\frac{i}{2}J_{HH}J_{HL}(M(t_1)-M(t_2))\{\hat{I}_{1z}(\hat{I}_{2x}\hat{L}_y - \hat{I}_{2y}\hat{L}_x) + \hat{I}_{2z}(\hat{I}_{1y}\hat{L}_x - \hat{I}_{1x}\hat{L}_y)\} + \frac{i}{2}J_{HH}J_{HL}(N(t_1)-N(t_2))\{\hat{I}_{1z}(\hat{I}_{2x}\hat{L}_x + \hat{I}_{2y}\hat{L}_y) - \hat{I}_{2z}(\hat{I}_{1x}\hat{L}_x + \hat{I}_{1y}\hat{L}_y)\}$$
$$+iJ_{HL}^2(M(t_1)N(t_2) - N(t_1)M(t_2))(\hat{I}_{1z} - \hat{L}_z)$$

$$[[\widetilde{\mathcal{H}}(t_1),\widetilde{\mathcal{H}}(t_2)],\widetilde{\mathcal{H}}(t_3)] = J_{HL}\left\{\left(\frac{(\Delta\omega_0)^2}{4} + \frac{J_{HH}^2}{2}\right)(M(t_1)-M(t_2)) + J_{HL}^2(M(t_1)N(t_2)-N(t_1)M(t_2))N(t_3)\right\}(\hat{I}_{1x}\hat{L}_x + \hat{I}_{1y}\hat{L}_y) \tag{S4}$$
$$+J_{HL}\left\{\left(\frac{(\Delta\omega_0)^2}{4} + \frac{J_{HH}^2}{2}\right)(N(t_2)-N(t_1)) + J_{HL}^2(M(t_1)N(t_2)-N(t_1)M(t_2))M(t_3)\right\}(\hat{I}_{1x}\hat{L}_y - \hat{I}_{1y}\hat{L}_x)$$
$$+\frac{\Delta\omega_0}{2}J_{HH}J_{HL}(M(t_1)-M(t_2))\left(\hat{I}_{1z}(\hat{I}_{2x}\hat{L}_x + \hat{I}_{2y}\hat{L}_y) - \hat{I}_{2z}(\hat{I}_{1x}\hat{L}_x + \hat{I}_{1y}\hat{L}_y)\right)$$
$$+\frac{\Delta\omega_0}{2}J_{HH}J_{HL}(N(t_1)-N(t_2))\left(\hat{I}_{1z}(\hat{I}_{2x}\hat{L}_y - \hat{I}_{2y}\hat{L}_x) + \hat{I}_{2z}(\hat{I}_{1y}\hat{L}_x - \hat{I}_{1x}\hat{L}_y)\right)$$
$$+\left(\frac{J_{HH}}{2} - \frac{J_{HL}}{4}\right)J_{HH}J_{HL}\{(M(t_2)-M(t_1))(\hat{I}_{2x}\hat{L}_x + \hat{I}_{2y}\hat{L}_y) + (N(t_2)-N(t_1))(\hat{I}_{2y}\hat{L}_x - \hat{I}_{2x}\hat{L}_y)\}$$
$$+\frac{J_{HH}(J_{HL})^2}{4}\{(M(t_2)-M(t_1))M(t_3) - (N(t_2)-N(t_1))N(t_3)\}(\hat{I}_{1x}\hat{I}_{2x} + \hat{I}_{1y}\hat{I}_{2y} + 2\hat{I}_{1z}\hat{I}_{2z} - 2\hat{I}_{2z}\hat{L}_z)$$
$$+\frac{J_{HH}(J_{HL})^2}{4}\{(N(t_2)-N(t_1))M(t_3) - (M(t_2)-M(t_1))N(t_3) + 2(M(t_1)N(t_2)-N(t_1)M(t_2))\}(\hat{I}_{1x}\hat{I}_{2y} - \hat{I}_{1y}\hat{I}_{2x})$$

However, to visualize the effect of the higher order average Hamiltonian, we plot how each matrix element varies with the pulse period T in Figure S1. The pulse period T should not be too long in most SABRE[3-6] systems with reasonable exchange rate (usually $\le 50s^{-1}$), otherwise the SABRE complex only experiences a constant high magnetic field during its lifetime. Therefore, we are only interested in pulse periods that are smaller or comparable to the lifetime of SABRE complex. For the sake of conciseness, here we only study one highly symmetric pulse, square pulse, and one less symmetric pulse--ramp. In addition, we only display the result of one subspace which is displayed in Figure S1; the other one has similar behavior. Since we only calculated the first and second order approximation, all the curves refer to the subspace of $\widetilde{\mathcal{H}}^{(1)} + \widetilde{\mathcal{H}}^{(2)}$. In the zero-order case, only the non-secular matric elements which connect the $\alpha_L$ states and $\beta_L$ states are altered, and the resulting $M$s and $N$s are identical. However, for the higher orders, diagonal elements are tuned as well, which is caused by the term $\hat{I}_{1z} - \hat{L}_z$ in the first order average Hamiltonian and the term $2\hat{I}_{1z}\hat{I}_{2z} - 2\hat{I}_{2z}\hat{L}_z$ in the second order (see equation (3) and (4)). Moreover, the $M$s and $N$s diverge in the higher order approximation. However, it is obvious that higher order average Hamiltonian are too small to rise obvious function (Figure S1). In conclusion, the zero order Magnus expansion is indeed a reliable approximation of the whole average Hamiltonian.

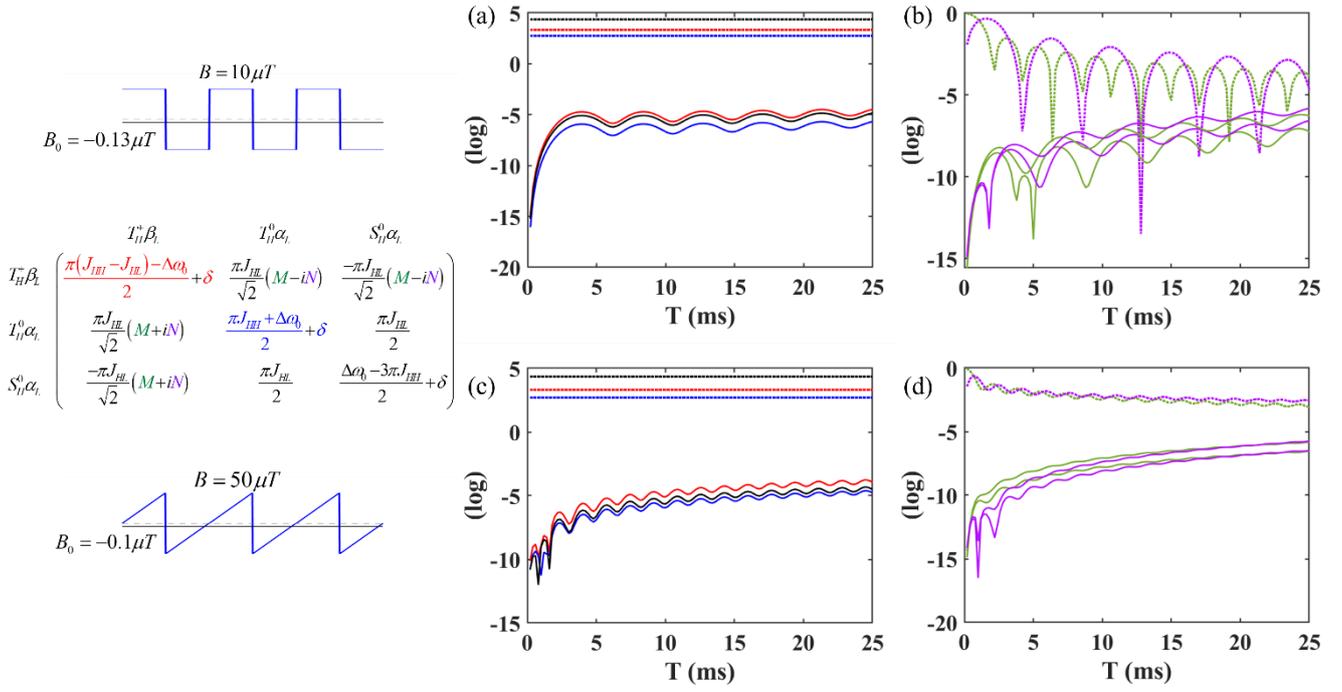

**Figure S1.** Plot of matrix elements in the effective Hamiltonian for a square pulse sequence (top) and ramp pulse sequence (bottom). The left and the right subplot column indicate how the three diagonal elements (left) and the adjustable coefficents M and N (right) change with the pulse period T. In all cases the dotted lines represent the zero order results while the full curves correspond to all higher orders (full numerical solution minus zero order). Because the higher order terms are almost always at least two orders of magnitude smaller than their zero-order counterparts, they are neglected in our theoretical analysis.

## M and N of a symmetric triangle pulse

In the article, we conclude that for an oscillating pulse which is symmetric both about the center of each repeating interval and the center of each half cycle, its corresponding $M_0$ and $N_0$ inevitably vanish together. Here, we provide a further verification by calculating one more example, a symmetric triangle pulse. Figure S2 shows how the curves of $M_0$, $N_0$ and $\sqrt{M_0^2 + N_0^2}$ of a symmetric triangle pulse vary as a function of $\theta$ (the rotation angle in half a period), respectively. The offset field is fixed at $-0.13\mu T$. When $\theta = (2n+1)\pi$, $M_0 = 0$, and when $\theta = 2n\pi$, $N_0 = 0$. Somewhere between $(2n-1)\pi$ and $2n\pi$, $M_0$ and $N_0$ vanish together. All of the above features agree with the analysis and proof given in the theory section of the article. Not that this triangle pulse gives $M_0$ and $N_0$ behaviors which are quite different from those of the Ramp pulse and nearly the same as those of the sine wave and square pulse, which indicates that the symmetry plays the key role.

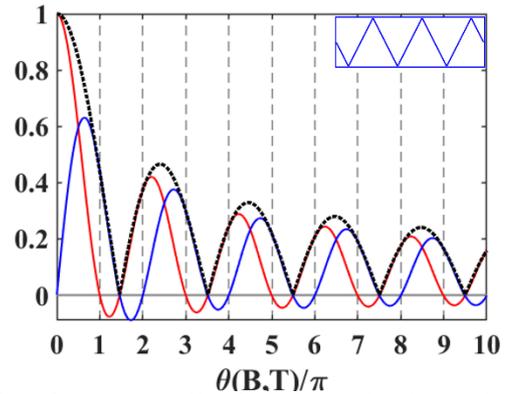

**Figure S2.** A depiction of how $M_0$ (red curve), $N_0$ (blue curve) and $\sqrt{M_0^2 + N_0^2}$ (black dotted curve) of the symmetric triangle pulse vary as a function of the rotation angle in half a period, respectively. We maintain the offset field at $-0.13\mu T$.

## Validity of Level Anti-crossings

Level anti-crossings[7,8] are a useful tool in many spectroscopic applications. For the last decade, they have been widely used in SABRE system to predict or explain the optimal magnetic field used in SABRE experiments. In the article we show that LAC works properly for our oscillating pulse SABRE-SHEATH[5, 6] even though it does not for CW SABRE-SHEATH[11]. Here, to support our conclusion we introduce a new model – a AA'B system with the coupling between the two hydrides being positive, $J_{HH} = 8 \text{Hz}$, while $J_{HL} = -25 \text{Hz}$ stays unchanged. A square pulse with amplitude $B = 10 \mu T$ is applied the spin system. Figure S3(a) depicts how polarization varies with both the offset field and the pulse period, in which the optimal offset field is about $\pm 0.38 \mu T$. Figure S3(b) displays the LACs of one $3 \times 3$ subspace with pulse period being fixed at the optimal value, 3.6ms, and the offset field changing from 0 to $0.4 \mu T$. The circled LAC at $B_0 \sim 0.37 \mu T$ agrees greatly with the optimal offset field. Therefore, the validity of LACs in Oscillating Pulse SABRE-SHEATH is verified again.

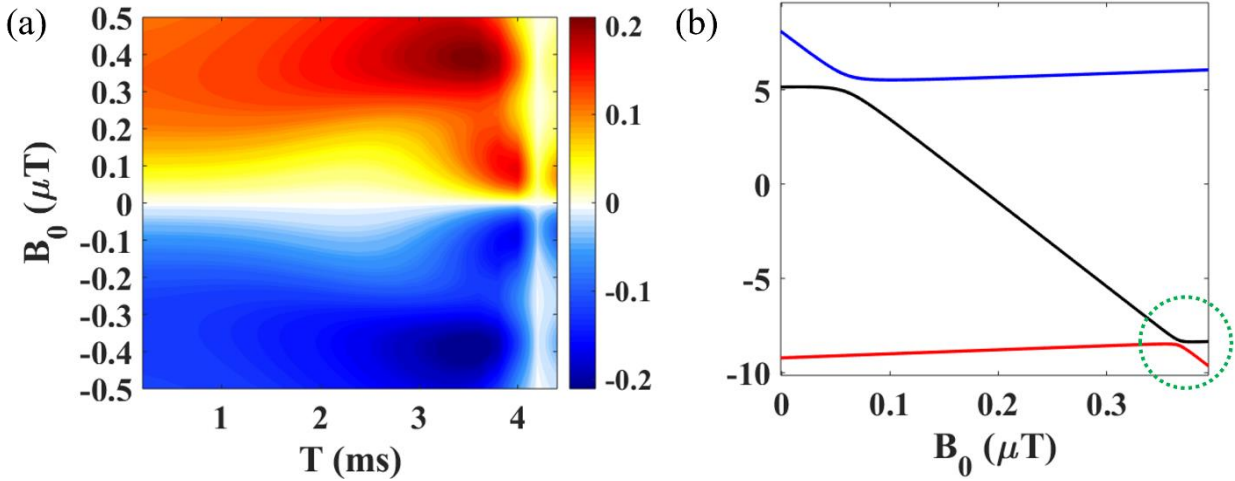

**Figure S3**. Experimental optimum agrees with the prediciton of LAC. The square pulse we use has a stable amplitude $10 \mu T$ while changing offset and pulse period. (a) displays polarizaiton as a funtion of both offset and pulse period. The optimal offset and the optimal pulse period are $B_0 = \pm 0.38 \mu T$, $T = 3.6 ms$. (b) is the corresponding LACs of one of the $3 \times 3$ subspaces. The circle LAC occurs near $B_0 \sim 0.37 \mu T$.

## Zero-order Average Hamiltonian of AA'BB' system

In the paper we demonstrated that the oscillating pulse makes the coupling between hydrides and target nuclei are adjustable in the case of 3-spin system AA'B. Here we extend this conclusion to symmetric AA'BB' system. The original full Hamiltonian of the 4-spin system experiencing an unbalanced squared pulse is

$$\widetilde{\mathcal{H}}_{prime}(t) = -\big(B_0 + B(t)\big)\big(\gamma_H(\hat{I}_{1z} + \hat{I}_{2z}) + \gamma_L(\hat{L}_{1z} + \hat{L}_{2z})\big) + 2\pi J_{HH}(\vec{I}_1 \cdot \vec{I}_2) + 2\pi J_{HL}(\vec{I}_1 \cdot \vec{L}_1) + 2\pi J_{HL}(\vec{I}_2 \cdot \vec{L}_2) \tag{S5}$$

In the same way as the 3-spin case, we rearrange the formula in the following

$$\widetilde{\mathcal{H}}_{rearranged}(t) = -\big(B_0 + B(t)\big)\gamma_H(\hat{I}_{1z} + \hat{I}_{2z} + \hat{L}_{1z} + \hat{L}_{2z}) + \big(\Delta\omega_0 + \Delta\omega(t)\big)(\hat{L}_{1z} + \hat{L}_{2z}) + 2\pi\{J_{HH}(\vec{I}_1 \cdot \vec{I}_2) + J_{HL}(\vec{I}_1 \cdot \vec{L}_1) + J_{HL}(\vec{I}_2 \cdot \vec{L}_2)\} \tag{S6}$$

in which $\Delta\omega_0 = B_0(\gamma_H - \gamma_L)$, and $\Delta\omega(t) = B(t)(\gamma_H - \gamma_L)$. Similarly, simplify the Hamiltonian by taking out the first term which commutes with the rest of the Hamiltonian.

$$\widetilde{\mathcal{H}}(t) = \big(\Delta\omega_0 + \Delta\omega(t)\big)(\hat{L}_{1z} + \hat{L}_{2z}) + 2\pi\{J_{HH}(\vec{I}_1 \cdot \vec{I}_2) + J_{HL}(\vec{I}_1 \cdot \vec{L}_1) + J_{HL}(\vec{I}_2 \cdot \vec{L}_2)\} \tag{S7}$$

Transfer this Hamiltonian to the toggling frame $U(t) = \exp\big(-i(\hat{L}_{1z} + \hat{L}_{2z})\int_0^t \Delta\omega(t')\,dt'\big)$

$$\begin{aligned}\widetilde{\mathcal{H}} &= U\{\Delta\omega_0(\hat{L}_{1z} + \hat{L}_{2z}) + 2\pi J_{HH}(\vec{I}_1 \cdot \vec{I}_2) + 2\pi J_{HL}(\vec{I}_1 \cdot \vec{L}_1) + 2\pi J_{HL}(\vec{I}_2 \cdot \vec{L}_2)\}U^\dagger \\ &= \Delta\omega_0(\hat{L}_{1z} + \hat{L}_{2z}) + 2\pi J_{HH}(\vec{I}_1 \cdot \vec{I}_2) + 2\pi J_{HL}\big(\hat{I}_{1z}\hat{L}_{1z} + M(t)(\hat{I}_{1x}\hat{L}_{1x} + \hat{I}_{1y}\hat{L}_{1y}) + N(t)(\hat{I}_{1x}\hat{L}_{1y} - \hat{I}_{1y}\hat{L}_{1x})\big) \\ &\quad + 2\pi J_{HL}\big(\hat{I}_{2z}\hat{L}_{2z} + M(t)(\hat{I}_{2x}\hat{L}_{2x} + \hat{I}_{2y}\hat{L}_{2y}) + N(t)(\hat{I}_{2x}\hat{L}_{2y} - \hat{I}_{2y}\hat{L}_{2x})\big)\end{aligned} \tag{S8}$$

Since there are two target nuclei in every SABRE complex, and each one couples with one of the two hydrides, these two couplings are adjusted identically by the oscillation pulse. The time dependent coefficients are $M(t) = \cos\big(\int_0^t \Delta\omega(t')\,dt'\big)$, and $N(t) = \sin\big(\int_0^t \Delta\omega(t')\,dt'\big)$, which are the same with the 3-spin case. Therefore, the zero-order average Hamiltonian is easy to work out.

$$\hat{\mathcal{H}}^{(0)} = \Delta\omega_0(\hat{L}_{1z} + \hat{L}_{2z}) + 2\pi J_{HH}(\vec{I}_1 \cdot \vec{I}_2) + 2\pi J_{HL}\left(\hat{I}_{1z}\hat{L}_{1z} + M_0(\hat{I}_{1x}\hat{L}_{1x} + \hat{I}_{1y}\hat{L}_{1y}) + N_0(\hat{I}_{1x}\hat{L}_{1y} - \hat{I}_{1y}\hat{L}_{1x})\right) \quad \text{(S9)}$$
$$+ 2\pi J_{HL}\left(\hat{I}_{2z}\hat{L}_{2z} + M_0(\hat{I}_{2x}\hat{L}_{2x} + \hat{I}_{2y}\hat{L}_{2y}) + N_0(\hat{I}_{2x}\hat{L}_{2y} - \hat{I}_{2y}\hat{L}_{2x})\right)$$

where $M_0 = \frac{\sin(\Delta\omega T/2)}{\Delta\omega T/2}$ and $N_0 = \frac{1-\cos(\Delta\omega T/2)}{\Delta\omega T/2}$. If $\Delta\omega T/2 = 2n\pi$, the coupling between hydride and the target nulcei vanish. Therefore, the resulting polarization is zero, as what is shown in Figure S4, along the vertical line of $T = 4.3ms$ the offset field $B_0$ and the pulse period T are scanned to figure out the optimal condition which is $B_0 = -0.16\mu T$, and $T = 4ms$, while the pulse amplitude B is pinned at $10\mu T$. Compared with the 3-spin system, the optimal pulse offset shifted a little bit which is caused by the different spin construction of the SABRE complex, but the optimal pulse period stays the same since $J_{HL}$ is tuned identically in both 3-spin and 4-spin cases. The subspaces of the average Hamiltonian associated with spin transfer are displayed in equation (S13). The basis used here is a singlet-triplet basis for both the AA' pair and the BB' pair. The couplings between the states of the target spins with respect to zero spin angular momentum projection ($S_L^0, T_L^0$) and states with a nonzero projection are adjustable. This fact is significant becasue the population transfer between these states, which is directly affected by the coupling strength, generates polarization, while the coupling between states $S_L^0$ and $T_L^0$ stays unchanged, and the coupling between states $T_L^+$ and $T_L^-$ is always zero.

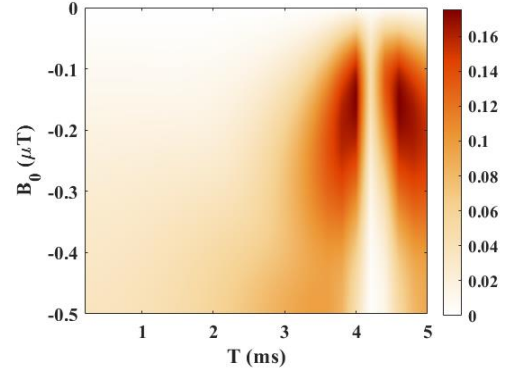

*Figure S4.* Polarization of a AA'BB' system under an unbalanced square pulse varies as the pulse period and the offset field. We hold the pulse amplitude, $B = 10\mu T$, and the accordingly optimal condition is $B_0 = -0.16\mu T$, and $T = 4ms$. Simulation parameters: coupling strength $J_{HH} = -8Hz$, $J_{HL} = -25Hz$, exchange rates $k_L = 24s^{-1}$, $k_H = 8s^{-1}$, [catalyst]:[ligand]=1:10.

$$\begin{array}{c}
\begin{array}{cc} T_H^+ S_L^0 & S_H^0 T_+^0 \end{array} \\
\begin{array}{c} T_H^+ S_L^0 \\ S_H^0 T_+^0 \end{array} \left( \begin{array}{cc} \frac{\pi J_{HH}}{2} & \pi J_{HL}(M_0 - iN_0) \\ \pi J_{HL}(M_0 + iN_0) & -\frac{3\pi J_{HH}}{2} - \Delta\omega_0 \end{array} \right) \\[1em]
\begin{array}{cc} T_H^- S_L^0 & S_H^0 T_L^- \end{array} \\
\begin{array}{c} T_H^- S_L^0 \\ S_H^0 T_L^- \end{array} \left( \begin{array}{cc} \frac{\pi J_{HH}}{2} & \pi J_{HL}(M_0 + iN_0) \\ \pi J_{HL}(M_0 - iN_0) & -\frac{3\pi J_{HH}}{2} + \Delta\omega_0 \end{array} \right) \\[1em]
\begin{array}{cccc} T_H^+ T_L^- & T_H^0 T_L^0 & S_H^0 S_L^0 & T_H^- T_L^+ \end{array} \\
\begin{array}{c} T_H^+ T_L^- \\ T_H^0 T_L^0 \\ S_H^0 S_L^0 \\ T_H^- T_L^+ \end{array} \left( \begin{array}{cccc} \frac{\pi(J_{HH} - 2J_{HL})}{2} + \Delta\omega_0 & \pi J_{HL}(M_0 - iN_0) & -\pi J_{HL}(M_0 - iN_0) & 0 \\ \pi J_{HL}(M_0 + iN_0) & \frac{\pi J_{HH}}{2} & \pi J_{HL} & \pi J_{HL}(M_0 - iN_0) \\ -\pi J_{HL}(M_0 + iN_0) & \pi J_{HL} & -\frac{3\pi J_{HH}}{2} & -\pi J_{HL}(M_0 - iN_0) \\ 0 & \pi J_{HL}(M_0 + iN_0) & -\pi J_{HL}(M_0 + iN_0) & \frac{\pi(J_{HH} - 2J_{HL})}{2} - \Delta\omega_0 \end{array} \right)
\end{array} \quad \text{(S10)}$$

## Robustness to variations of exchange rate

In the article we have already confirmed that the 3-spin system with negative $J_{HH}$ is robust to variations of exchange rate. In this section we demonstrate with DMEx[12] simulation method that the positive $J_{HH}$ case and 4-spin systems also have strong robustness to variations in exchange rate. Likewise, we use a square pulse sequence with $B = 10\mu T$ as an example. The optimal magnetic field of the AA'B system with $J_{HH} = 8Hz$ is $\pm 0.4\mu T$, while for the AA'BB' system with $J_{HH} = -8Hz$ the optimal filed is $\pm 0.11\mu T$. In Figure S5, we study 4 different cases with $k_L$ changing from $1s^{-1}$ to $100s^{-1}$, but the optimal field does not obviously shift. The optimal conditions do not shift a lot in the low exchange cases. As the exchange rate of the substrate goes up, the optimal condition shifts in the direction that the offset increases while the pulse period decreases. However, the overall shift is mild. All the feature are consistent with the 3-spin case shown in the article (Figure 8).

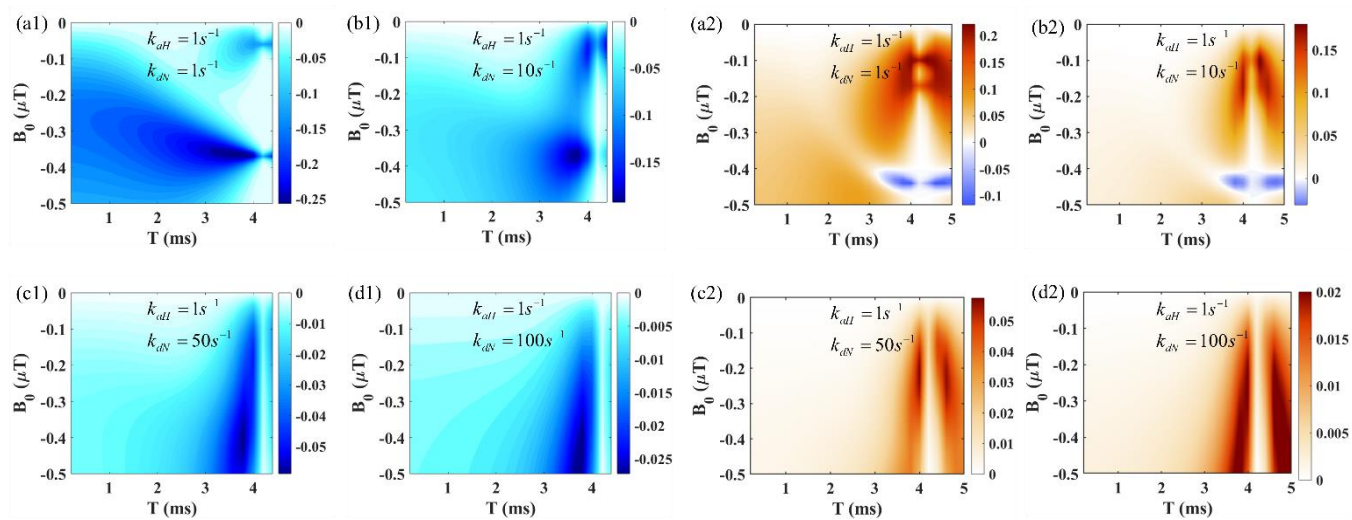

*Figure S5.* Optimal polarization conditions with different substrate exchange rates. The left four subplots refer to an AA'B spin system with a positive hydride coupling, $J_{HH} = 8Hz$, while the right four refer to an AA'BB' spin system with a negative hydride coupling, $J_{HH} = -8Hz$. (a1) The optimal field is -0.37µT, and the optimal period is 3.8ms. (b1) The optimal field is -0.37µT, and the optimal period is 3.8ms. (c1) The optimal field is -0.4µT, and the optimal period is 3.8ms. (d1) The optimal field is -0.44µT, and the optimal period is 3.8ms. (a2) The optimal field is -0.11µT, and the optimal period is 4.0ms. (b2) The optimal field is -0.17µT, and the optimal period is 4.0ms. (c2) The optimal field is -0.21µT, and the optimal period is 4.0ms. (d2) The optimal field is -0.5µT, and the optimal period is 3.8ms. The subplots have inconsistent color scales in order to show off more detail. Simulation parameters: coupling strength $J_{HL} = -25Hz$, [catalyst]:[ligand]=1:10.